\documentclass[letter,11pt]{article}

\usepackage{latexsym}
\usepackage{amsmath}
\usepackage[dvips]{graphicx}
\usepackage{graphpap}
\usepackage{pspicture}

\title{Naturally heavy superpartners and a Little Higgs}

\author{
         Tuhin Roy\, \thanks{\tt tuhin@physics.bu.edu}\ \
        and Martin Schmaltz\,\thanks{\tt schmaltz@physics.bu.edu} \ \ \\ 
        \\ 
        \small \sl \ Physics Department, Boston University, 
                Boston, MA  02215\\ \\ 
       }

\newcommand{\eqn}[1]{\label{eq:#1}}
\newcommand{\refeq}[1]{(\ref{eq:#1})}
\newcommand{\eq}{eq.~\refeq}

\newcommand{\Eq}{Eq.~\refeq}

\newcommand{\beq}{\begin{eqnarray}}
\newcommand{\eeq}{\end{eqnarray}}
\newcommand{\gsim}{\stackrel{>}{_\sim}}
\newcommand{\lsim}{\stackrel{<}{_\sim}}

\newcommand{\de}{\delta}

\newcommand{\un}{\mathrm{U}(1)}
\newcommand{\vl}{\langle}
\newcommand{\vr}{\rangle}
\newcommand{\tev}{\,\mathrm{TeV}}
\newcommand{\gev}{\,\mathrm{GeV}}
\newcommand{\mrm}{\mathrm}

\newcommand{\su}[1]{{\rm SU}(#1)}

\begin{document}

\baselineskip=17pt 
\pagestyle{plain}

\begin{titlepage} 
\vskip-.4in 
\maketitle 
\begin{picture}(0,0)(0,0) 
\put(360,200){BUHEP-05-13}
\end{picture}

\begin{abstract} 
\vskip.4in 
We construct an extension of the MSSM in which scalar superpartners can  
naturally be as heavy as 1 TeV. In the MSSM, the most significant fine tuning
stems from the logarithmically enhanced top-stop loop contribution to the
soft Higgs mass. We combine supersymmetry with the ``simplest little
Higgs" to render this loop finite, thereby  removing the large logarithm
even in models in which superpartner masses are generated at high scales such
as in supergravity.
Our model predicts an extended Higgs sector, superpartner masses near a
TeV and little Higgs partners at a few TeV. 

\end{abstract} 
\thispagestyle{empty} 
\setcounter{page}{0} 
\end{titlepage}

\section{Introduction}

In the Standard Model (SM) the Higgs mass receives quadratically divergent
quantum contributions. This indicates that the SM is only a
low-energy description of a more complete UV theory in which the
divergences are canceled by new particles. To avoid fine tuning
the masses of the new particles $\Lambda$ must be low enough
to cancel the quadratic divergence before it becomes
significantly bigger than the electroweak scale. For particles which couple
with coupling constants of order 1 to the Higgs this implies the
bound
\beq
\Lambda^2 \frac{1}{16 \pi^2}  \lsim m_{Higgs}^2  \ .
\eqn{quadratic}
\eeq
Numerically, one finds $\Lambda \lsim$ 2-5$\tev$,
within reach of precision electroweak measurements. No
significant deviations from the SM have been seen which implies
constraints on the couplings of these heavy particles.

In supersymmetry (SUSY), the new particles which cancel quadratic divergences
are the superpartners. 
An important feature of the minimal supersymmetric standard model (MSSM)
is that it is renormalizable. This means that no new physics
is required until extremely high scales, such as $M_{GUT}$ or $M_{Planck}$.
This is nice because it allows the MSSM to avoid problems with flavor
changing neutral currents (FCNCs) if we make the further assumption
of flavor universal soft SUSY breaking.
A problem with the MSSM however is that renormalization from all scales above
the superpartner masses ($M_{SUSY}$) enhances the one-loop contributions
to the Higgs mass parameter by a large logarithm and \eq{quadratic} is replaced by
\beq
M_{SUSY}^2\, \frac{1}{16 \pi^2}\ 
{\rm Log} (\frac{M_{Planck}^2}{M_{SUSY}^2})  \lsim m_{Higgs}^2 \ .
\eqn{logsensitivity}
\eeq
The large logarithm almost cancels the
$1/16 \pi^2$ and therefore naturalness requires superpartner masses
of order $m_{Higgs}$. Naively, this is a disaster for the MSSM
because it predicts a large number of superpartners at the weak scale
but none have been observed. The problem is greatly alleviated by 
R-parity~\cite{Farrar:1978xj}.
This symmetry implies that superpartners can only by
produced in pairs, and it eliminates all tree level contributions from SUSY to
precision electroweak observables. Nonetheless, the absence of any
evidence for superpartners is putting severe constraints on parameter space.

Another phenomenological problem of the MSSM is that in most of parameter
space the quartic Higgs coupling is too small and a physical Higgs
mass below the experimental bound is predicted. In the MSSM the only
way out of this problem is to increase the stop mass. This raises
the Higgs mass because the Higgs quartic receives contributions
from a top-stop loop proportional to $Log(m_{stop}/m_{top})$, and it also allows
larger corrections from A-terms.
A sufficiently large physical Higgs mass can be obtained with stop
masses larger than $1/2\tev$, but this comes at the cost of significant
fine tuning because of large contributions of the stops
to \eq{logsensitivity}. Various recent attempts to solve the fine
tuning problem in the MSSM may be found in
\cite{Batra:2003nj,Polonsky:2000rs, 
Birkedal:2004xi,Harnik:2003rs,Chacko:2005ra,Kitano:2005wc} 
and references therein.

The aim of this paper is to construct a supersymmetric model in which
superpartners 
can be heavy without destroying naturalness. The new ingredient is to introduce a
global symmetry into the Higgs sector of the MSSM which is
spontaneously broken at a few TeV. The lightest Higgs
doublet of the model arises as a pseudo-Nambu-Goldstone
boson (pNGB) in this symmetry breaking. Examples of these so-called
little-Higgs model building  in non-supersymmetric case may be found in 
\cite{Arkani-Hamed:2001nc,Arkani-Hamed:2002qx,Arkani-Hamed:2002qy,
Kaplan:2003uc,Schmaltz:2004de,Skiba:2003yf,Low:2002ws,Schmaltz:2002wx,
Chang:2003un,Chang:2003zn,Cheng:2003ju,Cheng:2004yc,Katz:2003sn}.
The combination of supersymmetry
and the global symmetry removes the logarithmic divergence in the soft
Higgs mass \eq{logsensitivity}. This allows superpartner masses to be
raised to near $4 \pi M_{weak}$ without significant tuning and
explains why no superpartners have been observed. It also increases
the loop contributions to the Higgs mass from the heavy stops. 

We construct our model by supersymmetrizing the ``simple group''
little Higgs \cite{Kaplan:2003uc,Schmaltz:2004de,Skiba:2003yf}.
The little Higgs mechanism and SUSY combine to
naturally generate a $4\pi$ hierarchy between the weak scale
and the ``partner'' scale (little Higgs partners and superpartners).
Unfortunately, this model with an $\su3_{weak}\times {\rm U}(1)$
gauge group and a minimal Higgs content predicts $\tan \beta =1$
which implies two problems: the usual tree level quartic Higgs
coupling vanishes and the predicted top Yukawa coupling is too small.
Our full model has a tree level quartic coupling for the Higgs which comes
from a supersymmetric version of the ``missing vev'' potential 
\cite{Kaplan:2003uc,Schmaltz:2005ky},
and we increase the top Yukawa couplings infrared quasi-fixed point
by splitting color $\su3$ into $\su3\times \su3$ at high energies. 

At the weak scale, our model reduces to the Standard Model
with a relatively light Higgs ($m_h \lsim 200\gev$).
Superpartners and an extended Higgs sector are near a TeV
while little Higgs partners are in the few TeV range. This places
the new physics out of reach of current accelerators but predicts
a rich physics program for the LHC.
In addition to the well-known SUSY signatures, the LHC might see signs
of the little Higgs quark partners ($T'$, $S'$, $D'$) and
the $\su3_{weak}$ gauge boson partners ($W'$ and $Z'$) which have
masses in the range of 1-5$\tev$.

The supersymmetrized ``simple group'' little Higgs model is
presented in Section 2. Section 3 contains the fully realistic
model with a tree level quartic and  Section 4 contains the phenomenology and
our conclusion.

\section{The Simplest Little SUSY}

The goal of this work is to construct a supersymmetric alternative
to the MSSM where the superpartners are parametrically heavier
than the electroweak symmetry breaking scale by a factor of $4\pi$ 
without fine tuning of the Higgs mass. 
In our model fine tuning is reduced relative to the MSSM,
and the model remains perturbative up to the Planck scale.

Our approach will be to construct a supersymmetric generalization of
a little Higgs theory where the quadratic divergence in the Higgs
mass is canceled because of a collectively broken approximate
global symmetry.
A new difficulty arises because of our
goal of constructing a UV complete theory which remains
well-defined all the way to $M_{Planck}$. The problem stems from
renormalization group running between $M_{Planck}$ and $M_{weak}$.
Collectively broken symmetries imposed at the Planck scale may
in fact be completely broken at low energies because perturbatively
small loop effects get enhanced by large logs. 
Consider for example, an exact global symmetry {\bf G} imposed at
$M_{Planck}$, but with a subgroup {\bf H} weakly gauged. As we renormalize
the theory down to low energies, wave function renormalization due
to {\bf H} interactions breaks the {\bf G} symmetry and -- because of the
large $\log(M_{Planck}/M_{weak})$ -- only {\bf H} remains as a
symmetry near the weak scale. This problem appears to be
generic to models in which a subgroup of a simple global symmetry 
is gauged (e.g. $[\su2]^2 \subset \su5$
in the Littlest Higgs or $[\su2]^2 \subset \su6$
in the $\su6/{\rm SP}(6)$ little Higgs). A naturally light Higgs mass 
in this framework can only occur if the soft masses are themselves
small enough. An alternative solution is to rely on a susy-breaking mechanism 
which generates soft masses at low energy (gauge mediation as an example). In
these models naturalness constraints the scale of susy breaking  to be around 
$100\tev$ \cite{Birkedal:2004xi,Chankowski:2004mq}. 

We therefore base our model on the Simplest Little Higgs 
\cite{Schmaltz:2004de} where the
global group $[\su3]^2$ is broken collectively by
gauging only the diagonal $\su3$. The nice feature 
here is that the log-divergent wave function renormalization
due to the $\su3$ gauge interactions automatically
preserves both $\su3$s.

The object of this Section is to construct a concrete
model where the above idea is realized.
We first construct the Higgs sector and show that it contains
a light electroweak doublet which we identify with the SM Higgs
doublet. We then show that it has all the necessary
interactions with matter as well as self couplings in order to break
electroweak symmetry. We show that radiative corrections to
the Higgs soft mass remain under control even when superpartner
masses are taken large. However, the model predicts $\tan \beta \simeq 1$ 
which implies that the top Yukawa coupling is too small. 
Another consequence  is that the Higgs
quartic coupling vanishes at tree level so that even
the large radiative correction due to the top-stop loop is not
sufficient to produce a Higgs mass above the current experimental limit. 

\subsection{Higgs sector and symmetries}

As in the Simplest Little Higgs we  start with the gauge group
$(\su3_c,\su3_w)_{\un_X}$ in the UV which is broken to the MSSM gauge
group at scale $f$. Third generation quarks and all leptons are
embedded in the following  $(\su3_c,\su3_w)_{\un_X}$ representations
\begin{eqnarray}
\Psi_{Q_3}&= (3,\bar{3})_{\frac{1}{3}} \qquad  \quad
	\Psi_L &= (1,\bar{3})_{- \frac{1}{3}}\nonumber \\
B^c &= (\bar{3},1)_{\frac{1}{3}} \qquad \quad
	E^c &= (1,1)_1  \nonumber \\
T^c_{1,2} &= (\bar{3},1)_{- \frac{2}{3}}  \qquad \quad
	N^c &= (1,1)_{0} \ .
\eqn{q.no-3gen}
\end{eqnarray}
The MSSM Higgs fields are embedded in fields $\Phi$ and their
adjoints $\bar{\Phi}$ with quantum numbers
\beq 
\Phi_1,\Phi_2 = (1,3)_{ \frac{1}{3}} \qquad \mrm{and} \qquad
\bar{\Phi}_1,\bar{\Phi}_2 = (1,\bar{3})_{-\frac{1}{3}} \ .
\eqn{q.no-phi}
\eeq
 Keeping in mind the symmetry breaking pattern  
$\su3_w \times \un_X \rightarrow \su2_w\times\un_Y$ we designate the
components of   $\Phi$ as
\beq
\Phi_i\equiv \left( \begin{array}{c}
                        H_i\\ S_i 
                     \end{array}\right) ; \hspace{3 mm}
\bar{\Phi}_i\equiv \left( \begin{array}{cc}
                        \bar{H}_i & \bar{S}_i 
                     \end{array}\right) \ .
\eeq 
Here $H_i$ and $\bar{H_i}$ are $ \su2_w$ doublets with hypercharge 
$+ \frac{1}{2}$ and $- \frac{1}{2}$, respectively. 
$\vl S_i\vr$ and  $\vl\bar{S}_i\vr$ 
break the gauge group $\su3_w \times \un_X$ down to $\su2_w\times\un_Y$.
The simplest superpotential which forces this symmetry breaking is 
\beq
W_{Higgs} = X_1(\bar{\Phi}_1 \Phi_1-f_1^2) + X_2(\bar{\Phi}_2\Phi_2-f_2^2),
\eqn{W-s-higgs}
\eeq  
where $X_1$ and $X_2$ are two gauge singlets.
Supersymmetry is broken softly as in the MSSM.
For reasons which have to do with the D-term potential and which are discussed at the
end of this subsection we assume that the
scalar components $\phi$ and $\bar \phi$ of the superfields $\Phi$ and
$\bar\Phi$ have identical soft masses
\beq
V_{soft} = \tilde{M}_1^2(|\phi_1|^2 + |\bar{\phi}_1|^2) + 
   \tilde{M}_2^2 (|\phi_2|^2 + |\bar{\phi}_2|^2) \ .
\eqn{v-soft}
\eeq
We will justify this assumption with an approximate $Z_2$
symmetry between  $\Phi$ and $\bar{\Phi}$ which may be the
remnant of a larger unifying symmetry at higher scales.

For simplicity, throughout this Section we work in the limit where the
soft masses are much smaller than the  $\su3_w$ breaking scale $f$.
The important result that the SM Higgs field is a
pseudo-Nambu-Goldstone boson (pNGB) is independent of
this limit. However, taking soft masses much smaller than $f$
has the advantage that the $\Phi_i$ fields may be expanded around
the supersymmetry preserving expectation values $f_i$, thereby keeping
intermediate formulas manifestly supersymmetric
\beq
\vl\phi_1\vr=\vl\bar{\phi}_1^\dag \vr =
         \left( \begin{array}{c}
       0 \\
       f_1
   \end{array}\right)&,&\hspace{3 mm}
\vl\phi_2\vr=\vl\bar{\phi}_2^\dag \vr = 
         \left( \begin{array}{c}
       0 \\
       f_2
   \end{array}\right) \ .
\eqn{vev1}
\eeq

We find that two linear combinations of doublets in the $\Phi$
and $\bar\Phi$s are eaten by the super-Higgs mechanism
\beq
H_{eat}\!\!\!&\equiv&\!\!\frac1f\Big[f_1 H_1 + f_2 H_2 \Big] \nonumber \\
\bar H_{eat}\!\!\!&\equiv&\!\!\frac1f\Big[f_1\bar{H}_1+f_2\bar{H}_2\Big] \ ,
\eqn{def-f-e/g}
\eeq
where $f \equiv \sqrt{f_1^2+f_2^2}$.
The two orthogonal combinations remain massless in the
supersymmetric limit
\beq
H_u  \!\!\!&\equiv& \!\!
         \frac{1}{f} \Big[f_2 H_1 - f_1 H_2 \Big] \nonumber \\
H_d \!\!\!& \equiv&\!\!
\frac{1}{f} \Big[f_2 \bar{H}_1 - f_1 \bar{H}_2 \Big]  \ .
\eqn{def-h-u/d}
\eeq
We now turn on soft SUSY breaking and expand the potential including
\eq{W-s-higgs}, \eq{v-soft}, and the D-term around the minima
in \eq{vev1}. After integrating out all scalars with masses of
order $f$ we arrive at an effective low energy potential
for the light doublets
\beq
V_{soft}^{eff} \approx (\tilde{M}_1^2 + \tilde{M}_2^2)\ |H_u-H_d^\dag|^2 
= 2  (\tilde{M}_1^2 + \tilde{M}_2^2)\, |\tilde{H}|^2  \ ,
\eqn{v-eff-soft}
\eeq
where
\beq
H &\equiv& \frac{1}{\sqrt{2}} \Big[H_u + H_d^\dag \Big]  \\
\tilde{H} &\equiv& \frac{1}{\sqrt{2}} \Big[H_u - H_d^\dag\Big] \ . 
\eqn{def-f-1/2}
\eeq
The doublet $H$ remains massless, it is protected
by the little Higgs mechanism. $\tilde{H}$ is very heavy
and does not contribute significantly to electroweak symmetry breaking.
The light field $H$ corresponds to the SM Higgs. Note that
a vacuum expectation value for $H$ implies equal vevs for $H_u$ and $H_d$, 
i.e. $\tan \beta = 1$.

For the remainder of this subsection we give an alternative 
derivation of the masslessness of the little Higgs.
Consider a similar theory in which two $\su3$ symmetries are gauged,
one acting on $\Phi_1,\bar \Phi_1$ and the other acting on
$\Phi_2,\bar \Phi_2$.
The two $\su3$ sectors are completely decoupled and can
be analyzed separately. Both have a single exact $\su3$ symmetry
which is spontaneously broken, leading to a set of NGBs which are
eaten by the Higgs mechanism. Note that SUSY is broken by the
soft masses and we expect no further massless scalar doublets.

Now recall our assumption of identical soft masses for
$\phi_i$ and $\bar\phi_i$. This implies identical vevs and therefore
a vanishing expectation value for the D-term 
(note, $D^a = \phi_1^\dagger t^a \phi_1-\bar\phi_1 t^a \bar\phi_1^\dagger
+ 1\leftrightarrow 2$).
and  only one linear combination of doublets 
obtains a mass from the D-term potential $(D^a)^2$. 
This combination is
$H_{eat}- \bar H_{eat}^\dag$. Note that this would not be
true if $D^8$ and $D^x$ had supersymmetry breaking expectation values, then
our little Higgs would also get a mass.

Finally, let us return to the theory with a single gauged $\su3$. The scalar
potential of this theory is identical except that there is now
a single D-term containing both sets of fields. But as before,
at quadratic order, this D-term does not depend on the linear
combinations $H_1+\bar H_1^\dagger$ and $H_2+\bar H_2^\dagger$.
The sum of these two correspond to the eaten doublet, the difference
remains massless, it is the little Higgs doublet $H$.
This argument can be extended to show that the little Higgs
also does not obtain a quartic self-coupling from
any terms in the tree-level potential either.

\subsection{Yukawa Couplings}
Yukawa couplings involving the top quark
are obtained from the superpotential 
\beq
W_{top} = Y_1 \Psi_{Q_3} \Phi_1 T^c_1 +  Y_2 \Psi_{Q_3} \Phi_2 T^c_2 \ . 
\eqn{w-top}
\eeq
where $Y_i$ are coupling constants of order 1.
After $\su3_w \times \un_X \rightarrow \su2_w\times\un_Y$ breaking,
the triplet $\Psi_{Q_3}$ is reducible. We denote the irreducible pieces as
\beq
\Psi_{Q_3} \equiv \left( \begin{array}{cc}
                       Q_3  & T 
                     \end{array}\right) \ .
\eeq
\Eq{w-top} gives mass to $T$
\beq
W_{top} \supset T (Y_1 f_1 T^c_1 + Y_2 f_2 T^c_2) \ .
\eqn{mass-T}
\eeq
The MSSM quark singlet $T^c$ is identified with the orthogonal combination of 
$T^c_1$ and $T^c_2$ which remains massless
\beq
T^c \equiv \frac{1}{\sqrt{Y_1^2 f_1^2 + Y_2^2 f_2^2}} 
                  \Big( Y_2 f_2 T^c_1 - Y_1 f_1 T^c_2 \Big) \ .
\label{def-T_c}
\eeq
The SM top quark Yukawa coupling is obtained from
\eq{w-top} by expanding to first order in the Higgs field
\beq
y_t = 
   \frac{Y_1 Y_2 f}{\surd{2} \sqrt{Y_1^2 f_1^2 + Y_2^2 f_2^2}} \ .
\eqn{y_t}
\eeq

Note that $\un_X$ and $\su3_w$ quantum numbers forbid a renormalizable
Yukawa coupling for the bottom quark. We therefore introduce an
additional triplet $\chi=(1,3)_{-\frac{2}{3}}$ and its adjoint
$\bar{\chi}=(1,\bar3)_{\frac{2}{3}}$ and write  
\beq
W_{\chi} = - M \chi\bar{\chi} 
      + Y_b \Psi_{Q_3}\chi B^c +\eta_1 \chi\Phi_1\Phi_2 
      + \eta_2 \bar{\chi} \bar{\Phi}_1 \bar{\Phi}_2 \ .  
\eqn{w-chi}    
\eeq
Here the $\su3_w$ indices in $\chi\Phi_1\Phi_2$ and in 
$\bar{\chi} \bar{\Phi}_1 \bar{\Phi}_2$ are contracted
anti-symmetrically with an epsilon tensor. After integrating
out $\chi$ at the scale $M\gg f$,
we obtain a Yukawa coupling for the bottom quark as well as
a $\mu$-term for the Higgses
\beq
W_{\chi}^{eff} & = &  
      \frac{Y_b\eta_2}{M} \Psi_{Q_3}(\bar{\Phi}_1 \bar{\Phi}_2) B^c +
      \frac{\eta_1\eta_2}{M} (\Phi_1 \Phi_2)(\bar{\Phi}_1\bar{\Phi}_2)  
                                  \nonumber \\
                          & \supset & 
       \frac{Y_b \eta_2 f}{M} Q_3 H_d B^c +
       \frac{\eta_1 \eta_2 f^2}{M} H_d H_u \ .
\eqn{w-eff-chi} 
\eeq
Note that since we used antisymmetric contractions, this part of
the potential does not change the vevs of $\mathcal O(f)$.
As in the MSSM, the $\mu$-term contributes to the Higgs
potential and gives masses to the Higgsinos.

\subsection{The Higgs mass and naturalness}

The standard model Higgs mass is related to its quartic coupling $\lambda$
by $m_h=\sqrt{\lambda} v$ where $v$ is the Higgs vev. Thus to determine
the Higgs mass, we need to know the quartic coupling. In our theory,
the Higgs is a pNGB and in the exact symmetry limit, the quartic
vanishes. 

A contribution to the quartic comes from the $(D^a)^2$-potential.
When the $D$-term vanishes this potential is insensitive
to supersymmetry breaking, and after integrating out the states
with masses of order $f$, the remaining potential is simply the MSSM
$D$-term 
\beq
\sum_a (H_u^\dag t^a H_u-H_d t^a H_d^\dag )^2 =
\sum_a (H^\dag t^a  \tilde{H} +\tilde{H}^\dag t^a H)^2 \ ,
\eqn{d-quartic}
\eeq
where $t^a$ are the $\su2 \times$U(1) generators. This looks
good, however there is a problem. Recall from the previous subsection
that $\tilde{H}$ has a mass of order the soft SUSY masses $\tilde M$ which are
large. Therefore the vev for $\tilde H$ vanishes to lowest order and
the D-term potential is flat in the $H$ direction.
This should be familiar from the MSSM where
the tree level quartic also vanishes for $\tan \beta =1$.

Below the superpartner masses, the Higgs quartic gets its usual loop
contributions~\cite{Haber:1990aw,Okada:1990vk,Ellis:1990nz}. 
The most important comes from the top quark and is cut off
at the stop mass
\beq
\de \lambda &\simeq&
    \frac{3y_t^4}{8 \pi^2} \ln \frac{\tilde{m}_t^2}{m_t^2} \ .
\label{v-one loop}
\eeq 
We see that the Higgs mass rises logarithmically with the stop mass, so that there
should be a critical stop mass for which this contribution is large enough
to lift the Higgs mass above the current bound
$m_h>114\gev$~\cite{Barate:2003sz}.
Unfortunately, the rise is very slow~\cite{Maloney:2004rc}
and even stop masses as large as $1\tev$ are not sufficient.
In the next Section we add new fields and interactions to the Higgs sector which
give a tree level contribution to the quartic coupling. We will see that 
Higgs masses between the experimental bound and about $200\gev$ are expected.

A closely related top-stop loop diagram gives a negative contribution to the soft
mass parameter of the Higgs and can trigger electroweak symmetry breaking
\beq
\de m^2 &\simeq& 
   - \frac{3y_t^2}{8 \pi^2} \tilde{m}_t^2 \ln \frac{f^2}{\tilde{m}_t^2} \ .
\eeq
Note that unlike in the MSSM the logarithm in this expression is cut off at the
$\su3$ breaking scale $f$, i.e. the logarithm is much smaller than in the
MSSM where we would have ln$({M^2_{Planck}}/{\tilde{m}_t^2})$
instead. This is the main accomplishment of the little Higgs mechanism in this 
context. The little Higgs partners render the stop loop finite above the scale
$f$ and therefore the Higgs mass is insensitive to the large scale $M_{Planck}$.
Plugging in numbers we see that stop masses as large as $1\tev$ do not
lead to significant fine-tuning.

\section{Extending Simplest Little SUSY }

The accomplishment of our model presented in the previous section is to
render the Higgs soft mass UV insensitive, that is to remove the large logarithm
which multiplies the Higgs soft mass in the MSSM and leads to excessive fine
tuning. The model also allows superpartner masses to be parametrically larger
than the electroweak scale, thus explaining why we have not seen any
signs of supersymmetry at colliders. In this section we aim to fix the three
technical problems which arose in the process. 
\begin{itemize}
\item  
  This model has a vanishing quartic Higgs coupling at tree level, requiring
  the Higgs mass to come from the top-stop loop. However, even for stop masses
  as large as $1\tev$, the Higgs mass appears to be too small. We therefore
  propose an extension of the simplest model with additional fields $\Phi_3$
  and $\bar \Phi_3$ which allow for the generation of a tree level quartic.
\item  
  Our little Higgs mechanism crucially depends on at least an approximate
  $Z_2$ charge conjugation in the Higgs sector. Unequal soft masses for 
  $\Phi$ and $\bar \Phi$ leads to a D-term of order the soft
  susy breaking and therefore also a Higgs soft mass of order $\tilde m^2$. A
  charge conjugation symmetry under which $\Phi \leftrightarrow \bar \Phi$
  is a way out of the problem. However, renormalization due to the top Yukawa
  coupling strongly breaks this  symmetry and -- in running from the Planck
  scale -- a large difference of order 1 is generated. To solve this problem
  we also make the top Yukawa coupling charge conjugation symmetric.
\item 
  Another problem is that the top Yukawa couplings $Y_1,
  Y_2$ in \eq{w-top} must be very large to ensure a sufficiently large top mass.
  When running them up into the UV they blow up at around $10^7\gev$.
  We show that an enlarged  color sector allows these couplings
  to stay perturbative all the way to the Planck scale.
\end{itemize}

\subsection{A tree level quartic}

In the previous section we pointed out that except for the small
couplings proportional to $\eta_1 \eta_2$ in
\eq{w-eff-chi} the superpotential preserves the two global $\su3$ symmetries
associated with $\Phi_1$ and  $\Phi_2$. As a result, the SM Higgs
doublet is a pNGB and receives neither a large mass nor a quartic.   
$W_{\chi}$ gives a weak scale mass and a tiny quartic.
Increasing the $\eta$'s in order to increase the quartic also raises the
mass term, thereby increasing fine tuning. In order to achieve natural electroweak
symmetry breaking with a heavy enough physical Higgs we need to
find a different source for a tree level quartic coupling.

Any single operator which gives a quartic by violating the two $\su3$s
also contributes to the soft mass term. To avoid this we employ the
little Higgs trick, collective symmetry breaking. We accomplish this by introducing
an extra field $\Phi_3$ which couples to $\Phi_1$ and $\Phi_2$ in two different terms
of the superpotential and hence breaks the two $\su3$s collectively. 
The new superpotential in the Higgs sector is
\beq
W_{Higgs}& = &X_1(\bar{\Phi}_1 \Phi_1-f_1^2) + X_2(\bar{\Phi}_2\Phi_2-f_2^2)
                                      + X_3 \bar{\Phi}_3 \Phi_3  \nonumber\\
  & + & a X_4\bar{\Phi}_1 \Phi_3 + b X_5 \bar{\Phi}_2 \Phi_3+ 
   c X_6 \bar{\Phi}_3 \Phi_1+ d X_7 \bar{\Phi}_3 \Phi_2+ W_s \ .
\label{W-higgs-phi3}
\eeq
$W_s$ is the part of the superpotential which involves only singlets. The
choice is not unique. We use the following simple form 
\beq
W_s = \frac{1}{3}X_3^3 - \frac{1}{2} m_3 X_3^2 
       +\frac{1}{3}X_4^3 - \frac{1}{2}m_4 X_4^2 
       +\delta X_5 X_6 X_7 \ .
\label{w-singlet}
\eeq

When all the vevs are turned on, the light Higgs doublet is rotated into 
$\bar{\Phi}_3$ and attains a quartic proportional to c and d. 
However it does not receive a large soft mass from this mixing as 
$\tilde M_3^2$ - the soft mass of $\bar{\Phi}_3$ - can be much smaller than
other soft masses. This is because $\Phi_3$ and $\bar{\Phi}_3$ do not couple
to colored particles and therefore the loop corrections to $\tilde M_3^2$ are
small. 

It is most transparent to analyze this theory when
$f^2 \gg m_3^2,m_4^2 $, and when all supersymmetric masses are larger than
soft masses 
because in this limit we can integrate out all the states associated with the
little Higgs mechanism in a supersymmetric fashion. In doing this, we find
the NMSSM~\cite{Nilles:1982dy,Ellis:1988er} with a 
distinctive prediction for superpartner masses at low energies.
However our results for the Higgs potential calculated in this section do not
depend on 
taking this limit, the model also works when some of the little Higgs states are
lighter than superpartners.

To begin, let us disregard $W_s$. Then $\su3_w$ is broken down
to $\su2_w$ at the scale $f$. One doublet is eaten by the heavy gauge fields
and one gets mass from the D-term. We end up with 4 light doublets and a
several singlets. Now as we turn on  $W_s$, the singlets $X_3,X_4$ and $X_5$
attain a non-zero vev. The superpotential in
eqs.(\ref{W-higgs-phi3},\ref{w-singlet}) gives mass 
to all but two doublets and singlets and we obtain a variant of the
NMSSM at low energies. 
\beq
W_h =  \rho S H_d H_u + \delta' S' S^2 \ .
\label{w-h-eff2}
\eeq
where we have defined  
\beq
H_u \!\!\!&\equiv& \!\!
         \frac{1}{f} \Big[f_2 H_1 - f_1 H_2 \Big] \nonumber \\
H_d \!\!\!& \equiv&\!\!   
          \frac{1}{f\sqrt{2\mu_4^2+\mu_3^2}}\Bigg[
            \mu_3 (f_2 \bar{H}_1 - f_1 \bar{H}_2) - 
            {\sqrt{2}\mu_4f} \bar{H}_3
         \Bigg] \ , 
\eeq    
and also parametrized 
\beq
\mu_3  & = & \mu_3 \vl X_3 \vr,\hspace{5 mm}
\sqrt{2} \mu_4 = a \vl X_4 \vr f/f_2 = - b \vl X_5 \vr f/f_1, \nonumber \\  
\rho & = &  \frac{cdf}{\sqrt{c^2f_1^2+d^2f_2^2}} 
          \frac{\sqrt{2}\mu_4}{\sqrt{2\mu_4^2+\mu_3^2}} \ .
\label{vev2}
\eeq     
Note that in this basis the soft terms are 
\beq
V_{soft} = (\tilde M_1^2 +\tilde M_2^2 ) \Bigg| H_u - 
       \frac{\mu_3}{\sqrt{2\mu_4^2+\mu_3^2}} H_d^\dag \Bigg|^2 + 
       \tilde M_3^2 \frac{2\mu_4^2}{2\mu_4^2+\mu_3^2} | H_d |^2 \ .
\eeq
Since $\tilde M_1^2,\tilde M_2^2 \gg \tilde M_3^2$, we identify the SM
Higgs doublet as 
\beq
H \equiv \frac{\mu_3}{\sqrt{2(\mu_4^2+\mu_3^2)}} H_u + 
         \frac{\sqrt{2\mu_4^2+\mu_3^2}} {\sqrt{2(\mu_4^2+\mu_3^2)}}H_d^\dag
   \ .    
\eeq
Defining the mixing angle $\theta_{34} \equiv \tan^{-1}(\mu_3/\mu_4)$
and using the Higgsino  mass 
$\mu \equiv \eta_1 \eta_2 \frac{f^2\mu_3}{M\sqrt{2\mu_4^2+\mu_3^2}}$
we can write the Higgs soft mass and quartic as
\beq
-m^2 &=&  |\mu |^2+ \tilde{M}_3^2 \cos^2 \theta_{34} \ , 
                         \nonumber \\
\lambda &=& \Bigg(\frac{|c|^2|d|^2 f^2}{8(|c|^2 f_1^2 + |d|^2 f_2^2)} \Bigg)
          \sin^2{2\theta_{34}} \ .
\label{v-h-theta34}
\eeq

\subsection{The $Z_2$ symmetry and anomalies}

At this point, we return to our assumption of equal soft masses for the
scalars $\phi$ and $\bar \phi$. We justified this assumption by showing that
the Higgs sector has a charge conjugation symmetry under which $\Phi
\leftrightarrow \bar \Phi$. 
However, the top Yukawa coupling breaks this symmetry and in running down from
the Planck scale a large asymmetry in the soft masses is generated.
To solve this problem we modify the top Yukawa coupling to make it charge conjugation
symmetric.

We add the multiplets $\Psi_{Q_4}=(3,3)_0$ and 
$\bar{\Psi}_{Q_4}=(\bar{3},\bar{3})_{0}$
as well as $B^c_2$ and $\bar{B}^c_2$, where $B^c_2$ has the same quantum numbers
as the MSSM $B^c$. These vector-like multiplets are assumed to have
masses near the scale 
$f$ so that the $Z_2$ symmetry of the Higgs-top sector is restored above $f$.
The charge conjugate of the top Yukawa coupling \eq{w-top} is then 
\beq
W_{Z_2} =  Y_1 \bar{\Phi}_1 \Psi_{Q_4} B^c + Y_2 \bar{\Phi}_2 \Psi_{Q_4} B^c_2   
\label{w-z2}
\eeq 
Of course, the $Z_2$ is violated by the masses for  $\Psi_{Q_4}$ and
$B^c_2$, but the  $\Phi-\bar{\Phi}$ soft mass splitting due to this symmetry 
breaking is small (loop suppressed and not log enhanced).

In extending the model to include first and second generation quarks we must cancel
the $SU(3)_w$ anomalies. This is most easily accomplished by embedding them as
\cite{Pisano:1991ee,Frampton:1992wt,Kong:2003tf,Schmaltz:2004de}
\beq
\Psi_{Q}^{(1,2)} = (3,3)_{0} \qquad  
\Psi_{Q}^{(3)} = (3,\bar{3})_{\frac{1}{3}}  \nonumber \\
D^{c(1,2)}_{1,2} = (\bar{3},1)_{\frac{1}{3}} \qquad
B^{c(3)} = (\bar{3},1)_{\frac{1}{3}}   \nonumber \\
U^{c(1,2)} = (\bar{3},1)_{- \frac{2}{3}}  \qquad
T^{c(3)}_{1,2} = (\bar{3},1)_{- \frac{2}{3}} 
\label{q.no-sm-a-free}
\eeq   

With this assignment of charges all anomalies vanish.
To generate Yukawa couplings for the up-type quarks in the first two generations
we couple them to $\bar{\chi}$ similarly to how we generated the bottom Yukawa.
The down-type quarks have renormalizable Yukawa couplings with $\bar{\Phi}$
\beq
W^{(1,2)}_{Yukawa} = Y_d^{(1,2)} \bar{\Phi}\Psi_{Q}^{(1,2)}D^{c(1,2)} +
   \frac{Y_u^{(1,2)}\eta_1}{M} \Psi_{Q}^{(1,2)} (\Phi_1\Phi_2) U^{c(1,2)}
\eeq
Note that one of the down type Yukawa couplings has to be sizable in order to give
large enough mass to the $D'$ and $S'$ partners. This
coupling cannot be smaller than $\sim 0.1$ but it should not be too large either
since these Yukawa couplings break the $\Phi \leftrightarrow \bar \Phi$
parity symmetry which is required to keep the Higgs naturally light. While
there is room for both conditions to be met, we expect that the $D'$ and $S'$ partners
are significantly lighter than $f\sim$ few TeV.

Note that embedding the first and second generation quarks differently into $SU(3)_w$ 
representations also violates the approximate flavor symmetries of the
MSSM and may lead to observable flavor changing effects.

\subsection{Avoiding a Landau pole for the top Yukawa coupling}

In the MSSM the electro-weak symmetry is broken by two multiplets $H_u$ and
$H_d$, with the relative size of the vacuum expectation values parameterized by
$\tan{\beta}$. Since the top Yukawa coupling $Y_T H_u Q_3T^c$ involves only
$H_u$, the coupling constant must be larger than the corresponding coupling in
the standard model $Y_T = \lambda_t /\sin{\beta}$. The 
soft masses of  $H_u$ and $H_d$ determine the relative size of the Higgs vevs
so that for equal soft masses one finds $\tan{\beta}=1$ and therefore 
$Y_T = \sqrt{2}\lambda_t$.
This is a problem in the MSSM because the top Yukawa coupling reaches a Landau pole
well before the unification scale.

The problem is even worse here.  Even before introducing the fields $\Phi_3$ and
$\bar \Phi_3$, the Higgs vev is evenly distributed between $\Phi$ and $\bar \Phi$
which corresponds to $\tan{\beta}=1$. However in order to generate a tree level
quartic coupling for the Higgs our model requires additional mixing with the
doublet in $\bar \Phi_3$ which further increases the required top Yukawa coupling.
To explore the divergence of the Yukawa couplings quantitatively, we analyze
the RGEs of $Y_1$ and  $Y_2$. Ignoring the small $\un_X$ gauge coupling
and all other Yukawa couplings they are
\beq
\frac{d}{dt}Y_1 &=& \frac{Y_1}{16\pi^2} \Big[
    7Y_1^2 + Y_2^2 - \frac{16}{3} (g_s^2+g_w^2) \Big] \nonumber \\
\frac{d}{dt}Y_2 &=& \frac{Y_2}{16\pi^2} \Big[
    7Y_2^2 + Y_1^2 - \frac{16}{3} (g_s^2+g_w^2) \Big]  \ .
\label{rge-y}
\eeq
As in the MSSM, there is an infrared quasi fixed point of the top Yukawa 
couplings at $Y_i^2 \sim \frac{2}{3} (g_s^2+g_w^2)$. 
Numerically, this fixed
point is almost identical to the MSSM fixed point at $\frac{1}{18} 
(16g_s^2+ 9g_w^2)$. Using $Y_1 = \sqrt{2}y_t$ at the TeV scale, we find that
$Y_1$ blows up near $10^7\gev$.

The root of the problem lies in the significant mixing between several Higgs
doublets, so that the strength of the top Yukawa coupling is diluted,
and larger values the primordial Yukawa couplings $Y_i$ are required.
To obtain these larger values without encountering Landau poles, we modify
the renormalization group equations by splitting the color group
$SU(3)_c$ into two. In the UV we take the third generation to
transform under $SU(3)_{c1}$ whereas the two light generations transform
under $SU(3)_{c2}$. At energies of order 5-10$\tev$ the two groups are broken
to the diagonal by expectation values for vector-like bi-fundamentals
$V$ and $\bar V$.

As can be seen from the matching relation
\beq
\frac{1}{g_s^2} = \frac{1}{g_{c1}^2} + \frac{1}{g_{c2}^2}
\label{def-g_c}
\eeq 
the coupling constant $g_{c1}$ which now replaces $g_s$ in the RGEs
can be significantly larger. In particular, since the one-loop
beta function for $g_{c1}^2$ vanishes exactly, $g_{c1}^2$ can
even be a few times larger than $g_{s}^2$ and yet remain perturbative.
Larger contributions from the gauge interactions in eq(\ref{rge-y}) in turn 
predict larger values of the fixed point Yukawa coupling.
In other words, much larger low energy values for $Y_1$ and $Y_2$
are now consistent with perturbativity all the way to the Planck scale.

Explicitly, the new (anomaly free) quantum numbers or quark and lepton
superfields under $(\su3_{c1}, \su3_{c2}, \su3_w)_{\un_X}$ are
\beq
\Psi_{Q}^{(1,2)} = (1,3,3)_{0} \qquad  
\Psi_{Q}^{(3)} = (3,1,\bar{3})_{\frac{1}{3}}  \qquad 
\Psi_{Q}^{(4)} = (3,1,3)_{0}\nonumber \\
D^{c(1,2)}_{1,2} = (1,\bar{3},1)_{\frac{1}{3}} \qquad
B^{c} = (\bar{3},1,1)_{\frac{1}{3}}   \qquad 
B^c_2 = (\bar{3},1,1)_{\frac{1}{3}} \nonumber \\
U^{c(1,2)} = (1,\bar{3},1)_{- \frac{2}{3}}  \qquad
T^{c}_{1,2} = (\bar{3},1,1)_{- \frac{2}{3}}  \qquad
V = (3,\bar{3},1)_{0} \ .
\label{q.no-uv}
\eeq   

In order to generate mixing between the third generation quarks and
those of the first and second generation ($V_{ub},V_{cb},V_{td}$ and $V_{ts}$)
additional mixing between the quarks with different color groups
is needed. This problem is easily solved with the help of the extra
colored multiplets which we introduced to implement the charge conjugation
$Z_2$ symmetry in the third generation. They allow us to write
renormalizable cross-generation Yukawa couplings.   
To be specific, 
\beq
W_{\Psi_{Q4}} = \xi_{1,2} \bar{\Psi}_{Q_4} V \Psi_{Q_{1,2}} 
      +  Y B^c \bar{\Phi}\Psi_{Q_4} + M_c\bar{\Psi}_{Q_4}\Psi_{Q_4}  
\label{w-q4-2}
\eeq 
and below $M_c$, $\Psi_{Q_4}$ and $\bar{\Psi}_{Q_4}$ can be integrated out
to yield the desired off-diagonal Yukawa couplings
\beq
W_{\Psi_{Q4}}^{eff} = -\Big(\frac{f_c}{M_c}\xi_{1,2}Y \Big) 
                        B^c \bar{\Phi}\Psi_{Q_{1,2}}\ \ .
\label{eq-new}
\eeq

\section{Phenomenology and Outlook}

\begin{figure}[t]
\begin{center}
\includegraphics[width=0.4\textwidth]{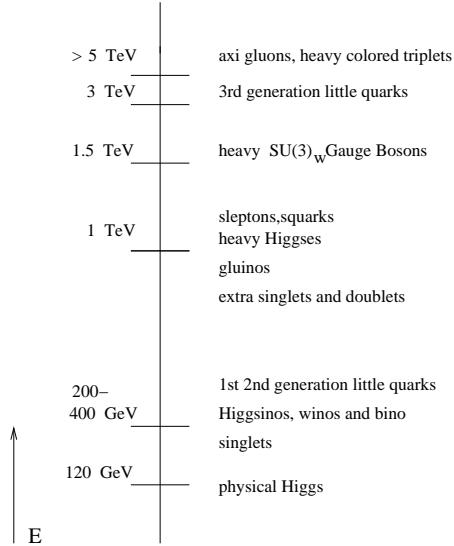}
\end{center}
\caption{A possible spectrum of the theory} 
\label{spectrum}
\end{figure}

This model offers very rich phenomenology. 
In addition to superpartners we expect to see new gauge bosons and fermions
from the little Higgs mechanism and new fields associated with the
doubled color sector.
There are four primary scales of phenomenology - 
the scale $f_c$ where the two different color gauge groups are broken to the SM
color group, the little Higgs scale $f$, the scale of soft scalar masses,
and the  gaugino/Higgsino masses. In Fig(\ref{spectrum}) we have depicted one
possible hierarchy.

At $f_c$ we find extra color gauge bosons, the axi-gluons. Their
mass squared is given by $\sim (g_{c1}^2+g_{c2}^2)f_c^2$ which is
heavier than $f_c^2$ because of the large values of the color gauge coupling
constants. Axi-gluons mediate $B_0-\bar{B}_0$ mixing at tree level which
implies a lower bound on the scale  $f_c$. Assuming that the new contribution
to the  $B_0-\bar{B}_0$ mixing amplitude is less than or equal to the SM
amplitude we find roughly $f_c \geq 5\tev$. The LHC will not be able to
produce such heavy axi-gluons with significant cross section.           

At the little Higgs scale $f$, we expect
to find new charged gauge bosons ($W'$s), neutral gauge bosons ($Z'$) as
well as new chiral multiplets which were needed in order to embed matter
particles into full $\su3_w$ multiplets ($T-\bar{T}$ for example). A lower 
bound on the scale $f$ can be obtained from contributions of the $Z'$ to precision
electroweak observables. Previous studies obtained bounds on the order of
$f\gsim 3 \tev$ which is the scale we have used for the explicit example in
the previous section. 
Naturalness implies that the little Higgs scale is near this lower limit, and
we expect that the LHC will be able to produce $W'$s, $Z'$
as well as little quarks. For details on the phenomenology of the Simplest Higgs see
\cite{Schmaltz:2005ky,Han:2004az,Han:2005dz,Han:2005ru,Kilian:2005xt,
Marandella:2005wd}.  

Further lower in the energy scale we expect a range of superpartners
along with an extra set of electroweak doublets from our
non-minimal Higgs sector. As is seen in  Fig(\ref{spectrum}) the scalar
superpartners are expected to be heavier than the gauginos. LHC will certainly 
produce $\tilde g \tilde g$, $\tilde q \tilde g$ and possibly even 
$\tilde q \tilde q$ pairs, which then decay via the usual susy cascades.
Chargino and neutralino production through valence
quark annihilation into weak bosons may also be significant. Since 
sleptons are even heavier than squarks, slepton pair production is rather
small.

\section{Note added}
While this manuscript was nearing completion we noticed reference 
\cite{Berezhiani:2005pb}
in which an extension of the Simplest Little Higgs to supersymmetry is considered.
This model is similar to our simplest little susy toy model in section 2 and
suffers from similar problems. In particular, the large contributions to the
soft mass of the little Higgs from the $T^8$ and $U(1)_x$ D terms have been
overlooked in  \cite{Berezhiani:2005pb}.

\section*{Acknowledgments}
We wish to thank David E. Kaplan and Takemichi Okui for helpful discussions
and the Aspen Center for Physics for it's hospitality. 
We acknowledge support from an Alfred P. Sloan Research Fellowship,
a U.S. Department of Energy Outstanding Junior Investigator Award
DE-FG02-91ER40676 and DOE grant DE-FG02-91ER40676.  

\bibliography{reference2}
\bibliographystyle{utcaps}

\end{document}